\documentstyle[12pt,WSPRL]{article}

\begin{document}

\begin{flushright}
July 1995\\IFP-719-UNC
\end{flushright}

\vspace*{1cm}
\centerline{\normalsize\bf THE LIGHT, THE HEAVY AND THE SUPERHEAVY}
\baselineskip=16pt
\vspace*{0.5cm}
\centerline{\normalsize\bf --- A NONABELIAN FLAVOR SYMMETRY FOR THE FULL
HIERARCHY\footnote{Talk given at International/Workshop on Particle
Phenomenology at IITAP, Ames, Iowa; May 1995.}}
\vspace*{1cm}

\centerline{\footnotesize OTTO C.W. KONG}
\baselineskip=13pt
\centerline{\footnotesize\it Institute of Field Physics, Department of Physics
and Astronomy,}
\baselineskip=12pt
\centerline{\footnotesize\it University of North Carolina, Chapel Hill, NC
27599-3255}
\centerline{\footnotesize E-mail: kong@physics.unc.edu}
\vspace*{2cm}

\centerline{ABATRACT}
\vspace*{0.5cm}
\normalsize
\baselineskip=20pt
We give a preliminary report of a new quark mass matrix model basing
on a $SU(5)\otimes SU(5)\otimes Q_{12}$ symmetry embedding into a fully
gauged $SU(5)\otimes SU(5)\otimes SU(2)$.~\cite{fk} The two $SU(5)$'s
 contain the standard SUSY $SU(5)$ as a diagonal subgroup, while the $Q_{12}$
or
$SU(2)$ is horizontal. Starting by assuming a judiciously-chosen set of
chiral supermultiplets, and a pattern of spontaneous symmetry breaking, we
obtain the low-energy chiral fermions together with a spectrum  of
 superheavy fermions at two different scales. The latter mediate
Froggatt-Nielsen
 tree graphs that give rise to a phenomenologically viable effective
 quark mass matrix texture. The model is the first example of a nontrivial
combination of supersymmetry without R-parity, gauged nonabelian horizontal
 symmetry and unification/anti-unification. It is expected to have some
very interesting features in SUSY-GUT phenomenology.

\newpage
\normalsize\baselineskip=15pt
\setcounter{footnote}{0}
\renewcommand{\thefootnote}{\alph{footnote}}
\section{The Light, the Heavy and the Superheavy}
The smallness of most of the quark mass and mixing parameters and the
strong hierarchy among them is one of the most interesting puzzle in particle
physics.  Flavor symmetry, especially a horizontal symmetry commuting with
 the Standard Model group or the GUT group, paired with the Froggatt-Nielsen
mechanism\cite{fn,lns}, provide a plausible explanation for the hierarchical
texture
pattern\cite{rrr} of the quark mass matrices. Following this popular approach,
here we present a new model using a discrete dicyclic subgroup\cite{Q2N},
$Q_{12}$,
of a gauged horizontal $SU(2)$.  The model is the first example of a nontrivial
combination of supersymmetry without R-parity, gauged nonabelian horizontal
 symmetry and unification/anti-unification\cite{ATG}.

We recall the hierarchy in the quark sector parameters given in powers of
$\lambda ~\sim ~0.22$, at around the GUT scale:\cite{lns,rrr}
\begin{center}
 $|V_{us}| \sim \lambda$, \ \ \ \  $|V_{cb}| \sim \lambda^2$,
\ \ \ \ $|V_{ub}| \sim \lambda^3 - \lambda^4$;
\end{center}
\begin{tabbing}
aaaaaaaaaaaaaaaaaa \= aaaaaaaaaaaaaaaaaaaaaaaa \= aaaaaaa \kill
 \> $m_u/m_c \sim \lambda^3 - \lambda^4$, \> $m_c/m_t \sim \lambda^3 -
\lambda^4$, \\
 \> $m_d/m_s \sim \lambda^2$, 		\> $m_s/m_d \sim \lambda^2$, \\
 \> $m_b/m_t \sim  \lambda^3$,
\> $m_t/\langle{H_u} \rangle  \sim 1$. \\
\end{tabbing}

To construct the light and heavy quark masses, the Froggatt-Nielsen
mechanism invokes a spectrum of superheavy fermions, which are essentially
in vector-like pairs, to communicate the effects of the flavor symmetry
breaking vevs to the low-energy chiral fermions. Integrating out the superheavy
particles then leaves us with effective quark mass terms  containing powers
of small parameters of the form $\frac{<S>}{M}$, say $\sim \lambda \ $ or
$\lambda^2$, where $<S>$ is a symmetry breaking vev and $M$ the superheavy
fermion mass allowed by the unbroken symmetry.

\section{The $SU(5)\otimes SU(5)\otimes SU(2)$ Model}
The model has a gauge symmetry given by $SU(5)\otimes SU(5)\otimes SU(2)$. For
the two $SU(5)$'s, they are the GUT groups for the third family
and the lighter two families respectively. The latter form a horizontal
doublet.

\begin{figure}

\setlength{\unitlength}{1.0cm}
\begin{center}
\begin{picture}(7,9)
\put(1.1,8){\framebox(4.8,0.7){$SU(5)\otimes SU(5)\otimes SU(2)$}}
\put(1.3,6){\framebox(4.2,0.7){$SU(5)\otimes SU(5)\otimes Q_{12}$}}
\put(2.5,4){\framebox(1.6,0.7){$SU(5)_D$}}
\put(1,1.5){\framebox(5,0.7){$SU(3)_c\otimes SU(2)_L\otimes U(1)_Y$}}
\put(6.2,8){$M_{SU(2)}$}
\put(6.2,6){$M_{Q_{12}}$}
\put(6.2,4){$M_{GUT}$}
\put(6.2,3){$M_{SUSY}$}
\put(6.2,4.5){$M_{0}$}
\put(6.2,1.5){$M_W$}
\put(3.3,7.8){\vector(0,-1){0.9}}
\put(3.3,5.8){\vector(0,-1){0.9}}
\put(3.3,3.8){\vector(0,-1){1.4}}
\put(3.3,1.3){\vector(0,-1){0.9}}

\end{picture}
\fcaption{The Symmetry Breaking Pattern of the Model}
\end{center}
\end{figure}

\subsection{Symmetry Breaking Pattern}
The two $SU(5)$'s then break into a diagonal $SU(5)$ which is
identified as the standard unification group, at energy scale $M_0$.
So below $M_0$ is the standard SUSY-GUT story. We assume that the
horizontal $SU(2)$ is broken to $Q_{12}$ at energy scale $M_{Q_{12}}$
with the latter subsequently totally broken at around $M_0$.
 The symmetry breaking pattern is summarised in Figure 1.

\subsection{Fermion Content}
The light, heavy and superheavy fermion contents of the model come from
the following list of $SU(5)\otimes SU(5)$ chiral supermultiplets:-

\begin{tabbing}
$SU(5)$ \= ---- \=SU(2) \= $Q_{12}$ 			\kill
\ \ \ \   {\em from} \> \> $SU(2)$ \> \ {\em to}  \hspace{0.5cm} $Q_{12}$ \\
$(10,1) $ \>--	\> $1$ \> $ \longrightarrow \hspace{0.5cm} 1(T)$ \\
	\>	\> $7$ \> $ \longrightarrow \hspace{0.5cm} 1^{'} + 2_2 + 2_4 + 2_6$ \\
	\>	\> $4$ \> $ \longrightarrow \hspace{0.5cm} 2_1 + 2_3$ \\

$(\bar{10},1) $ \>-- \> $7$ \> $ \longrightarrow \hspace{0.5cm} 1^{'} + 2_2 +
2_4 + 2_6$ \\
	\>	\> $4$ \> $ \longrightarrow \hspace{0.5cm} 2_1 + 2_3$ \\

$(\bar{5},1) $ \>--	\> $6$ \> $ \longrightarrow \hspace{0.5cm} 2_1 + 2_3 +
2_5(H_d/b)$ \\
	\>	\> $3$ \> $ \longrightarrow \hspace{0.5cm} 1^{'} + 2_2$ \\

$(5,1) $ \>--	\> $1$ \> $ \longrightarrow \hspace{0.5cm} 1(H_u)$ \\
	\>	\> $4$ \> $ \longrightarrow \hspace{0.5cm} 2_1 + 2_3$ \\
	\>	\> $3$ \> $ \longrightarrow \hspace{0.5cm} 1^{'} + 2_2$ \\

$(1,10) $ \>--	\> $2$ \> $ \longrightarrow \hspace{0.5cm} 2_1(Q)$ \\
	\>	\> $1$ \> $ \longrightarrow \hspace{0.5cm} 1$ \\

$(1,\bar{10}) $ \>-- \> $1$ \> $ \longrightarrow \hspace{0.5cm} 1$ \\

$(1,\bar{5}) $ \>--	\> $2$ \> $ \longrightarrow \hspace{0.5cm} 2_1(D)$ \\
(A summary of the $Q_{12}$ representations is given in the appendix.)
\end{tabbing}
\vspace*{0.3cm}

The vector-like fermion pairs have Dirac masses of order $M_{SU(2)}$
where $SU(2)$ is a good symmetry. At $M_{Q_{12}}$, the  two $(5,1)$ doublets,
$2_1$ and $2_3$ from the dimension $4$ $SU(2)$ representation married with
the corresponding $(\bar{5},1)$ doublets from the dimension $6$
$SU(2)$ representation to form Dirac fermions, leaving behind only the
labelled $Q_{12}$ singlets and doublets as GUT scale chiral particles.
The list of low energy chiral fermions is given by
$$\begin{array}{cccc}
\left. \begin{array}{c} \left( \begin{array}{c} u \\ d \end{array} \right)_{L}
\\
\left( \begin{array}{c} c \\ s \end{array} \right)_{L} \end{array}  \right\} &
Q(2_1)&
 \begin{array}{c} \left. \begin{array}{c} u_L^c\\c_L^c \end{array} \right\} \\
\left. \begin{array}{c} d_L^c\\s_L^c \end{array} \right\} \end{array}&
\begin{array}{c} Q(2_1)\\
D(2_1) \end{array} \\
\left( \begin{array}{c} t \\ b \end{array} \right)_{L}  & T(1)
&\begin{array}{c} t_L^c\\b_L^c
\end{array} & \begin{array}{c} T(1)\\H_d/b(2_5) \end{array} \\
\end{array}$$
and the correspondent leptonic partners of the GUT multiplets, and the
Higgsinos from $H_u$ --- $1$ and  $H_d/b$ --- $2_5$. Recall that in the
$SU(5)$ language, $H_u$ is a $5$, $Q$ and $T$ are $10$'s while $D$ and
the interesting $H_d/b$ are $\bar{5}$'s.

The horizontal doublet $H_d/b$ is definitely the most interesting element of
the model. It contains both the bottom-tau and the (down-sector) Higgs chiral
multiplets.
The group properties of the representation $2_5$ plays a very important role
in the model, as discussed below.

\subsection{Supersymmetry without R-parity}
The most interesting point to note about the assignment of the bottom-tau
and down-sector Higgs to a horizontal doublet is that it is incompatible
with the standard R-parity, which is put into MSSM by hand to avoid
unacceptable B and L violation. Here in our model, the group properties
of the horizontal symmetry gives this required matter parity feature\cite{MP}.

The only direct Yukawa couplings to Higgses are for the third family,
giving rank one quark mass matrices for both up and down sector at the first
order.
In terms of MSSM chiral superfields, the top get its mass from  the term
$\hat{q} \hat{h_1} \hat{u^c}$ contained in the $10.5.10$ coupling (Figure 2a).
The
bottom and the tau get their masses from the terms $\hat{q} \hat{h_2}
\hat{d^c}$
and $\hat{l} \hat{h_2} \hat{e^c}$ respectively, both contained in the
$10.\bar{5}.\bar{5}$ coupling (Figure3a). The latter however does not give rise
to
the dangerous terms $\hat{q} \hat{l} \hat{d^c}$,
$\hat{l} \hat{l} \hat{e^c}$, and $\hat{u^c} \hat{d^c} \hat{d^c}$. The
{\it secret} is in the $Q_{12}$ product $2_5 \times 2_5$ which contains the
singlet
$1$ only in the antisymmetric part, therefore always coupling the bottom-tau
part to the Higgs part but not to itself.

Moreover, no $\bar{5}.5$ term is allowed by the horizontal symmetry, hence
 both $\hat{l} \hat{h_1}$ and $\hat{h_2} \hat{h_1}$ are absent. In conclusion,
to the first order, the horizontal symmetry gives the required
 matter parity feature
 and evades the $\mu$-problem. Detailed properties of the model in
aspects like B and L violation, FCNC, and Higgsino masses and the role of
the horizontal symmetry in them is a very interesting question to be addressed.

\subsection{The Quark Mass Matrices}
All the above listed chiral supermultiplets are assumed to develop no vevs
above the $M_{GUT}$. We need then the following vector-like multiplet with
the specified vevs to generate the effective quark mass matrices. They are,
a  $<2_4>$ of $(1,1)$, a  $<2_1>$ of $(1,24)$, a $<2_1>$ and a $<2_2>$ of
$(\bar{10},10)$ coupled only to 2nd family in $Q$ or $D$, a $<2_2>$ of
$(5,\bar{5})$ coupled only to 2nd family in $Q$ or $D$, and  a $<2_1>$
of $(5,\bar{5})$ coupled only to 1st family in $Q$ or $D$.

The model is now completed. The lowest order effective quark mass term for
the up and down sector are given by the Froggatt-Nielsen tree graphs shown
in Figure 2 and 3 respectively. Taking
\begin{center}
$<2_k>/M_{Q_{12}} \sim \lambda$, \hspace{0.5cm}
$<2_k>/M_{SU(2)} \sim \lambda^2$; \hspace{0.5cm}
$<2_k> \sim M_0,$
\end{center}
we arrive at the quark mass textures
\begin{center}
\begin{math}
$$M_{u} \sim \left(\begin{array}{ccc}
\lambda^{8} & \lambda^{6} & 0 \\
\lambda^{6} & \lambda^{4} & 0 \\
0 & 0 & 1
\end{array}\right)$$,
\end{math}
\ \ \ \ \ \
\begin{math}
$$M_{d} \sim \left(\begin{array}{ccc}
0 & \lambda^{5} & \lambda^{5} \\
\lambda^{4} & \lambda^{3} & \lambda^{3} \\
0 & \lambda^{3} & 1
\end{array}\right)$$.
\end{math}
\end{center}

A comparison with the corresponding symmetric texture pattern
from reference 4, as given by
\begin{center}
\begin{math}
$$M_{u} \sim \left(\begin{array}{ccc}
0 & \lambda^{6} & 0 \\
\lambda^{6} & \lambda^{4} & 0 \\
0 & 0 & 1
\end{array}\right)$$,
\end{math}
\ \ \ \ \ \
\begin{math}
$$M_{d} \sim \left(\begin{array}{ccc}
0 & \lambda^{4} & 0  \\
\lambda^{4} & \lambda^{3} & \lambda^{3} \\
0 & \lambda^{3} & 1
\end{array}\right)$$.
\end{math}
\end{center}
is sufficiently convincing that it is
phenomenologically viable.

Possible alternative formulation of the model under a
$SU(5)\otimes U(1)\otimes SU(2)$\cite{fk} and comparison with the simple
$Q_{2N}$ models built under the same approach\cite{1} are skipped here.

\section{Acknowledgements}
The author want to thank his thesis adviser, P.H. Frampton, for helpful
guidance, constant encouragement and enjoyable collaboration. This work was
supported in part by the U.S. Department of
Energy under Grant DE-FG05-85ER-40219, Task B.\\

\section{Appendix on $Q_{2N}$ Representations\cite{Q2N,1}}
May be a dihedral group $D_N$ as the symmetry of a N-sided planar
polygon is more familiar to physicists.  The dicyclic group $Q_{2N}$
is a double cover of $D_N$, as the parent continous group $SU(2)$ being a
double cover of the rotation group $SO(3)$. $Q_{2N}$ is of order $4N$;
$Q_{12}$, the $N=6$ candidate in our model has order $24$.

Irreducible representations of $Q_{2N}$ are given by
 $4$ singlets $1, 1^{'}, 1^{''}, 1^{'''}$ and
$(N - 1)$ doublets $2_k$ (with $1 \leq k \leq (N - 1)$).
 Most important for our purposes are the
product formulae: -
\begin{equation}
1^{'} \times 1^{'} = 1
\end{equation}
\begin{equation}
1^{'}\times 2_k = 2_k
\end{equation}
\begin{equation}
2_k \times 2_l = 2_{(|k-l|)} + 2_{(min\{k+l,2N-k-l\})}
\end{equation}
where, in a generalized notation, $2_0 \equiv 1 + 1^{'}$
 and $2_N \equiv 1^{''} + 1^{'''}$.


\newpage

\begin{figure}


\setlength{\unitlength}{1.0cm}

\begin{picture}(25,19.7)

\thicklines

\put(15,19.4){$T - 1$}
\put(15,20){$H_u - 1$}
\put(15,18.8){$Q - 2_1$}

\put(1,18.5){\framebox(2.3,1){$(a)$: $(M_u)_{33}$}}
\put(7.08,17){\vector(1,0){1}}
\put(8.08,17){\line(1,0){0.84}}
\multiput(9,17)(0,0.3){9}{\line(0,1){0.25}}
\put(8.85,19.6){${\bf \times}$}
\put(9.08,17){\line(1,0){0.84}}
\put(10.92,17){\vector(-1,0){1}}
\put(9.2,19.6){$ \left \langle {H_u} \right \rangle $}
\put(11.2,16.9){$T(t)$}
\put(6,16.9){$T(t)$}

\put(1,13.5){\framebox(2.3,1){$(b)$: $(M_u)_{22}$}}
\put(7.08,12){\vector(1,0){1}}
\put(8.08,12){\line(1,0){0.84}}
\multiput(9,12)(0,0.3){9}{\line(0,1){0.25}}
\put(8.85,14.6){${\bf \times}$}
\put(9.08,12){\line(1,0){0.84}}
\put(10.92,12){\vector(-1,0){1}}
\multiput(11,12)(0,0.3){9}{\line(0,1){0.25}}
\put(10.85,14.6){${\bf \times}$}
\put(11.08,12){\line(1,0){0.84}}
\put(12.92,12){\vector(-1,0){0.85}}
\put(12.07,12){\vector(-1,0){0.15}}
\put(4,11.9){$Q (c)$}
\put(9.2,14.6){$ \left \langle {H_u} \right \rangle $}
\put(10.2,11.5){$2_2$}
\put(11.2,14.6){$ \left \langle {2_1} \right \rangle $}
\put(13.2,11.9){$Q (c)$}
\put(5.08,12){\vector(1,0){0.85}}
\put(5.93,12){\vector(1,0){0.15}}
\put(6.08,12){\line(1,0){0.84}}
\multiput(7,12)(0,0.3){9}{\line(0,1){0.25}}
\put(6.85,14.6){${\bf \times}$}
\put(7.2,14.6){$ \left \langle {2_1} \right \rangle $}
\put(7.5,11.5){$2_2$}

\put(1,8.5){\framebox(3.8,1){$(c)$: $(M_u)_{21}/(M_u)_{12}$}}
\put(7.08,7){\vector(1,0){1}}
\put(8.08,7){\line(1,0){0.84}}
\multiput(9,7)(0,0.3){9}{\line(0,1){0.25}}
\put(8.85,9.6){${\bf \times}$}
\put(9.08,7){\line(1,0){0.84}}
\put(10.92,7){\vector(-1,0){1}}
\multiput(11,7)(0,0.3){9}{\line(0,1){0.25}}
\put(10.85,9.6){${\bf \times}$}
\put(11.08,7){\line(1,0){0.84}}
\put(12.92,7){\vector(-1,0){0.85}}
\put(12.07,7){\vector(-1,0){0.15}}
\put(4,6.9){$Q (c)$}
\put(9.2,9.6){$ \left \langle {H_u} \right \rangle $}
\put(10.2,6.5){$2_2$}
\put(11.2,9.6){$ \left \langle {2_2} \right \rangle $}

\put(5.08,7){\vector(1,0){0.85}}
\put(5.93,7){\vector(1,0){0.15}}
\put(6.08,7){\line(1,0){0.84}}
\multiput(7,7)(0,0.3){9}{\line(0,1){0.25}}
\put(6.85,9.6){${\bf \times}$}
\put(7.2,9.6){$ \left \langle {2_1} \right \rangle $}
\put(7.5,6.5){$2_2$}
\multiput(13,7)(0,0.3){9}{\line(0,1){0.25}}
\put(12.85,9.6){${\bf \times}$}
\put(13.08,7){\line(1,0){0.84}}
\put(14.92,7){\vector(-1,0){0.85}}
\put(14.07,7){\vector(-1,0){0.15}}
\put(15.2,6.9){$Q (u)$}
\put(12.2,6.5){$1$}
\put(13.2,9.6){$ \left \langle {2_1} \right \rangle $}

\put(1,3.5){\framebox(2.3,1){$(d)$: $(M_u)_{11}$}}
\put(7.08,2){\vector(1,0){1}}
\put(8.08,2){\line(1,0){0.84}}
\multiput(9,2)(0,0.3){9}{\line(0,1){0.25}}
\put(8.85,4.6){${\bf \times}$}
\put(9.08,2){\line(1,0){0.84}}
\put(10.92,2){\vector(-1,0){1}}
\multiput(11,2)(0,0.3){9}{\line(0,1){0.25}}
\put(10.85,4.6){${\bf \times}$}
\put(11.08,2){\line(1,0){0.84}}
\put(12.92,2){\vector(-1,0){0.85}}
\put(12.07,2){\vector(-1,0){0.15}}
\put(2,1.9){$Q (u)$}
\put(9.2,4.6){$ \left \langle {H_u} \right \rangle $}
\put(10.2,1.5){$2_2$}
\put(11.2,4.6){$ \left \langle {2_2} \right \rangle $}
\put(5.08,2){\vector(1,0){0.85}}
\put(5.93,2){\vector(1,0){0.15}}
\put(6.08,2){\line(1,0){0.84}}
\multiput(7,2)(0,0.3){9}{\line(0,1){0.25}}
\put(6.85,4.6){${\bf \times}$}
\put(3.08,2){\vector(1,0){0.85}}
\put(3.93,2){\vector(1,0){0.15}}
\put(4.08,2){\line(1,0){0.84}}
\multiput(5,2)(0,0.3){9}{\line(0,1){0.25}}
\put(4.85,4.6){${\bf \times}$}
\put(7.5,1.5){$2_2$}
\put(5.5,1.5){$1$}
\put(7.2,4.6){$ \left \langle {2_2} \right \rangle $}
\put(5.2,4.6){$ \left \langle {2_1} \right \rangle $}
\multiput(13,2)(0,0.3){9}{\line(0,1){0.25}}
\put(12.85,4.6){${\bf \times}$}
\put(13.08,2){\line(1,0){0.84}}
\put(14.92,2){\vector(-1,0){0.85}}
\put(14.07,2){\vector(-1,0){0.15}}
\put(15.2,1.9){$Q (u)$}
\put(12.2,1.5){$1$}
\put(13.2,4.6){$ \left \langle {2_1} \right \rangle $}

\end{picture}

\fcaption{Froggatt-Nielsen tree graphs for $M_u$.*}


\end{figure}


\newpage

\begin{figure}


\setlength{\unitlength}{1.0cm}

\begin{picture}(25,19.7)

\thicklines

\put(15,20){$T - 1$}
\put(15,19.4){$H_d/b - 2_5$}
\put(15,18.8){$Q - 2_1$}
\put(15,18.2){$D - 2_1$}

\put(1,18.5){\framebox(2.3,1){$(a)$: $(M_d)_{33}$}}
\put(7.08,17){\vector(1,0){1}}
\put(8.08,17){\line(1,0){0.84}}
\multiput(9,17)(0,0.3){9}{\line(0,1){0.25}}
\put(8.85,19.6){${\bf \times}$}
\put(9.08,17){\line(1,0){0.84}}
\put(10.92,17){\vector(-1,0){1}}
\put(9.2,19.6){$ \left \langle {H_d} \right \rangle $}
\put(11.2,16.9){$b$}
\put(6,16.9){$T(b)$}

\put(1,13.5){\framebox(2.3,1){$(b)$: $(M_d)_{32}$}}
\put(7.08,12){\vector(1,0){1}}
\put(8.08,12){\line(1,0){0.84}}
\multiput(9,12)(0,0.3){9}{\line(0,1){0.25}}
\put(8.85,14.6){${\bf \times}$}
\put(9.08,12){\line(1,0){0.84}}
\put(10.92,12){\vector(-1,0){1}}
\multiput(11,12)(0,0.3){9}{\line(0,1){0.25}}
\put(10.85,14.6){${\bf \times}$}
\put(11.08,12){\line(1,0){0.84}}
\put(12.92,12){\vector(-1,0){0.85}}
\put(12.07,12){\vector(-1,0){0.15}}
\put(4,11.9){$T(b)$}
\put(9.2,14.6){$ \left \langle {H_d} \right \rangle $}
\put(10.2,11.5){$2_1$}
\put(11.2,14.6){$ \left \langle {2_2} \right \rangle $}
\put(13.2,11.9){$D (s)$}
\put(5.08,12){\vector(1,0){1}}
\put(6.08,12){\line(1,0){0.84}}
\multiput(7,12)(0,0.3){9}{\line(0,1){0.25}}
\put(6.85,14.6){${\bf \times}$}
\put(7.2,14.6){$ \left \langle {2_4} \right \rangle $}
\put(7.5,11.5){$2_4$}

\put(1,8.5){\framebox(2.3,1){$(c)$: $(M_d)_{23}$}}
\put(7.08,7){\vector(1,0){1}}
\put(8.08,7){\line(1,0){0.84}}
\multiput(9,7)(0,0.3){9}{\line(0,1){0.25}}
\put(8.85,9.6){${\bf \times}$}
\put(9.08,7){\line(1,0){0.84}}
\put(10.92,7){\vector(-1,0){1}}
\multiput(11,7)(0,0.3){9}{\line(0,1){0.25}}
\put(10.85,9.6){${\bf \times}$}
\put(11.08,7){\line(1,0){0.84}}
\put(12.92,7){\vector(-1,0){1}}
\put(4,6.9){$Q (s)$}
\put(9.2,9.6){$ \left \langle {H_d} \right \rangle $}
\put(10.2,6.5){$2_3$}
\put(11.2,9.6){$ \left \langle {2_4} \right \rangle $}
\put(13.2,6.9){$b$}
\put(5.08,7){\vector(1,0){0.85}}
\put(5.93,7){\vector(1,0){0.15}}
\put(6.08,7){\line(1,0){0.84}}
\multiput(7,7)(0,0.3){9}{\line(0,1){0.25}}
\put(6.85,9.6){${\bf \times}$}
\put(7.2,9.6){$ \left \langle {2_1} \right \rangle $}
\put(7.5,6.5){$2_2$}

\put(1,3.5){\framebox(2.3,1){$(d)$: $(M_d)_{13}$}}
\put(9.08,2){\vector(1,0){1}}
\put(10.08,2){\line(1,0){0.84}}
\multiput(11,2)(0,0.3){9}{\line(0,1){0.25}}
\put(10.85,4.6){${\bf \times}$}
\put(11.08,2){\line(1,0){0.84}}
\put(12.92,2){\vector(-1,0){1}}
\multiput(13,2)(0,0.3){9}{\line(0,1){0.25}}
\put(12.85,4.6){${\bf \times}$}
\put(13.08,2){\line(1,0){0.84}}
\put(14.92,2){\vector(-1,0){1}}
\put(4,1.9){$Q (d)$}
\put(11.2,4.6){$ \left \langle {H_d} \right \rangle $}
\put(12.2,1.5){$2_3$}
\put(13.2,4.6){$ \left \langle {2_4} \right \rangle $}
\put(7.08,2){\vector(1,0){0.85}}
\put(7.93,2){\vector(1,0){0.15}}
\put(8.08,2){\line(1,0){0.84}}
\multiput(9,2)(0,0.3){9}{\line(0,1){0.25}}
\put(8.85,4.6){${\bf \times}$}
\put(5.08,2){\vector(1,0){0.85}}
\put(5.93,2){\vector(1,0){0.15}}
\put(6.08,2){\line(1,0){0.84}}
\multiput(7,2)(0,0.3){9}{\line(0,1){0.25}}
\put(6.85,4.6){${\bf \times}$}
\put(9.5,1.5){$2_2$}
\put(7.5,1.5){$1$}
\put(7.2,4.6){$ \left \langle {2_1} \right \rangle $}
\put(9.2,4.6){$ \left \langle {2_2} \right \rangle $}
\put(15.2,1.9){$b$}

\end{picture}

\fcaption{Froggatt-Nielsen tree graphs for $M_d$.*}


\end{figure}


\newpage

\addtocounter{figure}{-1}%

\begin{flushleft}
\begin{figure}


\setlength{\unitlength}{1.0cm}

\begin{picture}(25,14.7)

\thicklines

\put(1,13.5){\framebox(2.3,1){$(e)$: $(M_d)_{22}$}}
\put(5.08,12){\vector(1,0){0.85}}
\put(5.93,12){\vector(1,0){0.15}}
\put(8.08,12){\line(1,0){0.84}}
\multiput(9,12)(0,0.3){9}{\line(0,1){0.25}}
\put(8.85,14.6){${\bf \times}$}
\put(9.08,12){\line(1,0){0.84}}
\put(10.92,12){\vector(-1,0){1}}
\multiput(11,12)(0,0.3){9}{\line(0,1){0.25}}
\put(10.85,14.6){${\bf \times}$}
\put(11.08,12){\line(1,0){0.84}}
\put(12.92,12){\vector(-1,0){0.85}}
\put(12.07,12){\vector(-1,0){0.15}}
\put(4,11.9){$Q (s)$}
\put(9.2,14.6){$ \left \langle {H_d} \right \rangle $}
\put(10.2,11.5){$2_3$}
\put(11.2,14.6){$ \left \langle {2_2} \right \rangle $}
\put(13.2,11.9){$D (s)$}
\put(7.08,12){\vector(1,0){1}}
\put(6.08,12){\line(1,0){0.84}}
\multiput(7,12)(0,0.3){9}{\line(0,1){0.25}}
\put(6.85,14.6){${\bf \times}$}
\put(7.2,14.6){$ \left \langle {2_1} \right \rangle $}
\put(7.5,11.5){$2_2$}

\put(1,8.5){\framebox(2.3,1){$(f)$: $(M_d)_{21}$}}
\put(7.08,7){\vector(1,0){1}}
\put(8.08,7){\line(1,0){0.84}}
\multiput(9,7)(0,0.3){9}{\line(0,1){0.25}}
\put(8.85,9.6){${\bf \times}$}
\put(9.08,7){\line(1,0){0.84}}
\put(10.92,7){\vector(-1,0){1}}
\multiput(11,7)(0,0.3){9}{\line(0,1){0.25}}
\put(10.85,9.6){${\bf \times}$}
\put(11.08,7){\line(1,0){0.84}}
\put(12.92,7){\vector(-1,0){0.85}}
\put(12.07,7){\vector(-1,0){0.15}}
\put(4,6.9){$Q (s)$}
\put(9.2,9.6){$ \left \langle {H_d} \right \rangle $}
\put(10.2,6.5){$2_2$}
\put(11.2,9.6){$ \left \langle {2_1} \right \rangle $}
\put(13.2,6.9){$D (d)$}
\put(5.08,7){\vector(1,0){0.85}}
\put(5.93,7){\vector(1,0){0.15}}
\put(6.08,7){\line(1,0){0.84}}
\multiput(7,7)(0,0.3){9}{\line(0,1){0.25}}
\put(6.85,9.6){${\bf \times}$}
\put(7.2,9.6){$ \left \langle {2_2} \right \rangle $}
\put(7.5,6.5){$2_3$}

\put(1,3.5){\framebox(2.3,1){$(g)$: $(M_d)_{12}$}}
\put(9.08,2){\vector(1,0){1}}
\put(10.08,2){\line(1,0){0.84}}
\multiput(11,2)(0,0.3){9}{\line(0,1){0.25}}
\put(10.85,4.6){${\bf \times}$}
\put(11.08,2){\line(1,0){0.84}}
\put(12.92,2){\vector(-1,0){1}}
\multiput(13,2)(0,0.3){9}{\line(0,1){0.25}}
\put(12.85,4.6){${\bf \times}$}
\put(13.08,2){\line(1,0){0.84}}
\put(14.92,2){\vector(-1,0){0.85}}
\put(14.07,2){\vector(-1,0){0.15}}
\put(4,1.9){$Q (s)$}
\put(11.2,4.6){$ \left \langle {H_d} \right \rangle $}
\put(12.2,1.5){$2_3$}
\put(13.2,4.6){$ \left \langle {2_2} \right \rangle $}
\put(7.08,2){\vector(1,0){0.85}}
\put(7.93,2){\vector(1,0){0.15}}
\put(8.08,2){\line(1,0){0.84}}
\multiput(9,2)(0,0.3){9}{\line(0,1){0.25}}
\put(8.85,4.6){${\bf \times}$}
\put(5.08,2){\vector(1,0){0.85}}
\put(5.93,2){\vector(1,0){0.15}}
\put(6.08,2){\line(1,0){0.84}}
\multiput(7,2)(0,0.3){9}{\line(0,1){0.25}}
\put(6.85,4.6){${\bf \times}$}
\put(9.5,1.5){$2_2$}
\put(7.5,1.5){$1$}
\put(7.2,4.6){$ \left \langle {2_1} \right \rangle $}
\put(9.2,4.6){$ \left \langle {2_2} \right \rangle $}
\put(15.2,1.9){$D (d)$}

\end{picture}

\fcaption{Froggatt-Nielsen tree graphs for $M_d.$*}

\vspace*{0.7cm}

\hrule
\footnotesize * In both Figure 2 and 3, the superheavy fermions and scalar vevs
 are indicated by their representations in terms of the $Q_{12}$ symmetry,
while the low energy chiral particles are indicated by their label. All
vertical dash-lines represen

t scalar vevs. In Figure 2, all horizontal lines represent fermions; those
shown by  arrows are $(10,1)$'s while those shown by  double-arrows are
$(1,10)$'s under $SU(5) \otimes SU(5)$. Likewise in  Figure 3, except that
those left pointing arrows and do

uble-arrows (fermions to the right of $H_d$) are here $(\bar{5},1)$'s and
$(1,\bar{5})$'s respectively. The full representations for the scalars can then
be figured out easily.


\end{figure}

\end{flushleft}

\end{document}